\documentclass[12pt]{article}
\usepackage{amssymb}

\begin{document}

\author{E. Ahmed$^{1}$, A. S. Hegazi$^{1}$ and A. S. Elgazzar$^{2,3}$ \\
$^{1.}$Mathematics Department, Faculty of Science\\
35516 Mansoura, Egypt\\
$^{2.}$Mathematics Department, Faculty of Education\\
45111 El-Arish, Egypt\\
$^{3.}$Mathematics Department, Faculty of Science - Al-Jouf\\
King Saud University, Kingdom of Saudi Arabia}
\title{On Difference Equations Motivated by Modelling the Heart}
\date{}
\maketitle

\begin{abstract}
The analytical structure of some generalizations of the circle map
is given. Also a generalization of off centre reflection is
studied. The stability of Ito-Glass coupled map lattice is
studied.\newline
\end{abstract}

\section{Introduction}

Circle map is one of the subjects which is both interesting mathematically
and useful in applications [Glass et al 1989, Vandenmeer 2001]. It relates,
in a beautiful way, some pure mathematical problems and applications.

For example in [Glass et al 1989] it is shown that the simplest circle map $%
f(\theta )=\theta +\tau _{0}\;\mathrm{{mod}\;1}$, where $\tau _{0}$ is a
constant is related to a human heart disease (parasystol) where there are
two types of heart pulses. This is represented by a symbolic sequence of
2-symbols (say 0,1). Such sequence defines another one called the reduced
sequence where one starts with 1 and count the number of iterates till the
next 1 appears and so on. This reduced sequence has remarkable properties
with some relevance to number theory: i) The reduced sequence contains, at
most, three integers say $p,m,n$. ii) $p=m+n$. iii) $m$ or $n$ is odd. iv)
Only one of $p,m,n$ can succeed itself in the reduced sequence. These
results agree with clinical observations.

There are , however, some diseases e.g. tachyacardia which should be
represented by coupled map lattice [Ito and Glass 1992]. Therefore in this
paper we study some generalizations of the circle map analytically. The
paper is organized as follows: In Section 2, the definition of the circle
map is given, and some generalizations are introduced. Many generalizations
Ito-Glass coupled map that is more suitable for modelling some heart
diseases are introduced and studied in section 3. The conditions for the
stability of its equilibrium solutions are derived. Some conclusions are
summarized in section 4.

\section{Circle map and some generalizations}

In the following, some basic results about circle map will be reviewed and
generalized.\newline
\newline
\textbf{Definition 1:} A circle map is a map $f:S^{1}\rightarrow S^{1}$,
where $S^{1}$ is the unit circle. A map $F:R\rightarrow R$ is the lift of
the circle map $f$ if
\begin{equation}
\pi \circ f=F\circ \pi,
\end{equation}
where $R$ is the real numbers and $F\circ \pi$ means the superposition of
the two maps $F$ and $\pi$, and
\begin{equation}
\pi :R\rightarrow S^{1},\;\pi (t)=\exp (2\pi it)\;\forall \;t\in R.
\end{equation}
\newline
\newline
In this case $F$ (hence $f$) is of degree one if
\begin{equation}
F(x+1)=F(x)+1.
\end{equation}
\newline
\newline
\textbf{Definition 2:} \newline
(a) The rotation number of a circle map $f$ is defined by
\begin{equation}
\rho (f,x)=\lim_{n\rightarrow \infty }\frac{F^{n}(x)-x}{n},\;F^{n}=F\circ
F^{n-1}.
\end{equation}
\newline
\newline
(b) If $f$ is differentiable then the Lyapunov exponent of $f$ is defined by
\begin{equation}
\lambda =\lim_{n\rightarrow \infty }\frac{1}{n}\sum_{j=0}^{n-1}\ln \left|
f^{\prime }(x_{j})\right| ,
\end{equation}
where
\begin{equation}
x_{j+1}=f(x_{j}),\;j=0,1,2,....
\end{equation}
\newline
\newline
In general $\rho (f,x)$, if it exists, depends on the initial point and
forms a closed interval [Ito 1981].

The following is a known result yet for the sake of completeness we present
a proof:\newline
\newline
\textbf{Proposition 1: }(a) If $F(x)$ is monotonically increasing and
continuous then $\rho (f,x)$ is independent of $x$.\newline
\newline
(b)\ $\rho (f^{q})=q\rho (f)\;$for all$\;$positive integers $q$.\newline
\newline
\textbf{Proof.} Since $F$ is monotonically increasing and continuous (hence
bounded) then $\lim_{n\rightarrow \infty }(F^{n}(x)-x)/n$ exists.

To show that it is independent of $x$ assume, without loss of generality, $%
0\leq x\leq y\leq 1$. Since $F$ is monotonically increasing then there is a
positive integer $k$ such that
\[
F^{k}(x)\leq F(y)\leq F^{k+1}(x),
\]
hence
\[
F^{n+k-1}(x)\leq F^{n}(y)\leq F^{n+k}(x).
\]
Taking the limit as $n\rightarrow \infty ,$ one gets
\[
\lim_{n\rightarrow \infty }F^{n}(x)\;=\lim_{n\rightarrow \infty }F^{n}(y).
\]
Recalling that $-1\leq x-y\leq 0,$ then
\[
\lim_{n\rightarrow \infty }\frac{F^{n}(x)-x}{n}\;=\lim_{n\rightarrow \infty }%
\frac{F^{n}(y)-y}{n}.\;
\]
To prove (b) use
\[
\lim_{n\rightarrow \infty }\frac{F^{nq}(x)-x}{n}=q\lim_{qn\rightarrow \infty
}\frac{F^{nq}(x)-x}{qn}.\;
\]
This completes the proof. $\blacksquare $\newline
\newline
\textbf{Lemma 1 }[Boyland 1986]: Let $f:S^{1}\rightarrow S^{1}$ be degree 1
continuous circle map, then $\forall p,q$ positive integers, $p/q\in \rho
(f,x)$ if and only if $f$ has a $p/q$ periodic point i.e. $F^{q}(x)=x+p.$%
\newline
\newline
\textbf{Definition 3}: A circle map $g$ is in class $A$ if $g$ is
continuous, degree one and if its lift $G$ is piecewise strictly monotone
with precisely two critical points $(g^{\prime }(x)=0)$ in $(0,1)$ one
maximum at $m_{2}$ and one minimum at $m_{1}$ and $m_{1}>m_{2}.$\newline
\newline
\textbf{Definition 4}: Let
\begin{equation}
F=x+w+bp(x),\;\;\;p:R\rightarrow R,\;p\mathrm{\;is\;continuous}%
,\;\;p(x+1)=p(x),\;\;0\leq w<1,
\end{equation}
such that

\begin{enumerate}
\item[(a)]  $p(0)=p(1)=0$.

\item[(b)]  If $b>1$ then $F$ is in class $A$.

\item[(c)]  If $b<1$ then $F$ is a homeomorphism.

\item[(d)]  $p(x)$ is the restriction to the real line of a complex entire
map.
\end{enumerate}

\textbf{Theorem 1} [Boyland 1986]: Let $F$ be as in Definition 4 and let $f $
be the corresponding circle map then the boundaries of the regions in the $%
b-w$ plane where $p (f)$ is periodic (these regions are typically called
Arnold tongues) are continuous curves.\newline
\newline
\textbf{Corollary 1:} Let
\begin{equation}
F(x)=x+w+\frac{r}{2\pi }\sin (2\pi x)+\frac{r^{2}}{4\pi }\sin (4\pi x),
\end{equation}
then Arnold tongues for $F(x)$ are continuous curves for $r<0.5$.\newline
\newline
\textbf{Proof.} Comparing Eq. (7) and Eq. (8), then
\begin{equation}
p(x)=\sin (2\pi x)+\frac{r}{2}\sin (4\pi x),\;\;b=\frac{r}{2\pi }.
\end{equation}
It is direct to see that $p(x)$ in Eq. (9) satisfies all the conditions for
Definition 4, hence using Theorem 1, the corollary is proved. $\blacksquare $%
\newline

There are several ways to generalize the above results. One way is to use
circle maps other than the standard sine circle map
\begin{equation}
f(x)=x+w+\frac{a}{2\pi }\sin (2\pi x),
\end{equation}
as has been done in Eq. (8) and will be done in the following. Alternatively
maps with degree other than one will be studied. This direction was
pioneered by Keener [Keener 1980] and extended by Glass and his group
[Belair and Glass 1985, Bub and Glass 1995]. This direction will be
discussed further later on.\newline
\newline
\textbf{Definition 5} [Au 2001]: The off centre reflection map $%
f:S^{1}\rightarrow S^{1}$ is defined by
\begin{equation}
f=x+w-2\{\tan ^{-1}[\sin \frac{2\pi x}{\cos 2\pi x-r}]-x\},\;0\leq
r<1,\;0\leq w<1.
\end{equation}
\newline

The following proposition was briefly proposed in [Au 2001] paper. Here the
full proof is given:\newline
\newline
\textbf{Proposition 2:} The off centre reflection map $f(x)$ in Eq. (11) can
be written as
\begin{equation}
f(x)=x+w-2\sum_{k=1}^{\infty }\frac{r^{k}}{k}\sin (2\pi kx).
\end{equation}
\newline
\newline
\textbf{Proof.} The third term of $f(x)$ in Eq. (11) is an odd function
hence it is representable as Fourier sine series hence
\[
f(x)=x+w-2\sum_{k=1}^{\infty }b_{k}(r)\sin (2\pi kx),
\]
where
\[
\frac{\partial b_{k}}{\partial r}=\frac{2}{\pi }\int_{0}^{\pi }\frac{\cos
(k-1)u-\cos (k+1)u}{1-2r\cos u+r^{2}}du.
\]
Define
\[
I_{k}=\frac{2}{\pi }\int_{0}^{\pi }\frac{\cos (ku)}{1-2r\cos u+r^{2}}du,
\]
and use
\[
\frac{1}{1-2r\cos u+r^{2}}=(1-r\exp (iu))^{-1}(1-r\exp (-iu))^{-1}.
\]
Expanding in powers of $r$, one gets
\[
I_{k}=\sum_{n=0}^{\infty }r^{2n+k}=\frac{r^{k}}{1-r^{2}},
\]
thus one gets
\[
b_{k}=\frac{r^{k}}{k},
\]
which completes the proof. $\blacksquare $\newline
\newline
\textbf{Corollary 2:} The function in Eq. (12) is phase of the restriction
on the unit circle of the complex function
\begin{equation}
h(z)=\exp (iw)\frac{z^{2}(1-rz)}{z-r},\;0\leq r<1.
\end{equation}
\newline

The proof is by taking the logarithm of both sides and expanding in powers
of $r$.

It is straightforward to derive the following result [Au 2001]:\newline
\newline
\textbf{Proposition 3:}

\begin{enumerate}
\item[(a)]  For $0\leq r\leq 1/3,$ the off centre reflection map $f$ in Eq.
(12) is monotonic hence its rotation number is a single point independent of
the initial value.

\item[(b)]  There is a saddle node bifurcation at
\begin{equation}
\left| w-\frac{1}{2}\right| =\frac{a_{r}}{\pi },\;\cos a_{r}=r.
\end{equation}

\item[(c)]  The off centre reflection map $f,$ Eq. (12) has a 2-cycle if $w=0
$\ or\ $w=1/2$ and a period doubling bifurcation at $r=1/\sqrt{5}$ to a
4-cycle.
\end{enumerate}

Now we consider maps of degree $M(M\neq 1)$ i.e.
\begin{equation}
f(x+1)=f(x)+M.
\end{equation}
\newline
\newline
\textbf{Definition 6 }[Belair and Glass 1985]: The rotation number for a
degree $M$ map is defined by
\begin{equation}
\rho (f,x)=\lim_{n\rightarrow \infty }\frac{1}{n}%
\sum_{j=0}^{n-1}[F(x_{j})-x_{j}],
\end{equation}
where $F$ is the lift of $f$.\newline
\newline
\textbf{Proposition 4}: [Belair and Glass 1989]

\begin{enumerate}
\item[(a)]  Let
\begin{equation}
f(x,w)=g(x,w)+w,
\end{equation}
be a circle map with degree $M$ then $\forall w$ real number there is at
least $(\sum_{j=0}^{n-1}M^{j-1})$ different values between $w,\;w+1$ for
which $x$ is a periodic point of $f^{n}.$

\item[(b)]  If $g$ is strictly monotonic, $g=g(x,b)$, $b$ is a real
parameter, then there is a continuous curve $w(b)$ along which $x$ is a
periodic point with period $n$.
\end{enumerate}

\textbf{Definition 7} [Keener 1980]\textbf{:} Let $F$ be the lift of a
circle map with degree $M$. Following Keener [Keener 1980] the following
properties are defined:

\begin{enumerate}
\item[ ]  $F$ satisfies property FI if $F(x)$ is piecewise continuously
differentiable function defined on the interval [0,1] with a single
discontinuity at $x=\theta \;\in (0,1).$

\item[ ]  $F$ satisfies property FII\ if it satisfies FI and
\begin{equation}
x\neq \theta \;\Rightarrow \;0<\frac{\mathrm{d}F}{\mathrm{d}x}<\infty .
\end{equation}

\item[ ]  $F$ satisfies property FIII\ if it satisfies FII and
\begin{equation}
\lim_{x\rightarrow \theta ^{-}}F(x)=1,\;\;\lim_{x\rightarrow \theta
^{+}}F(x)=0\;,\;\;\;\;\;F(\theta )=0\;\mathrm{or}\;1\;\theta \in (0,1).
\end{equation}

where $F$ has a single discontinuity at $x=\theta \;\in (0,1).$

\item[ ]  $F$ satisfies property FIV\ if it satisfies FIII and $F(0)>F(1)$
(non-overlapping case).

\item[ ]  $F$ satisfies property FV\ if it satisfies FIII and $F(1)>F(0)$
(overlapping case).
\end{enumerate}

\textbf{Definition 8:} The system (6) is chaotic in the standard sense if
its Lyapunov exponent, Eq. (5) is positive. It is chaotic in the sense of
Keener if its rotation number, Eq. (4) depends on $x$.\newline

In the following we present two results. The first is about the rotation
number of a generalization of the truncated off centre reflection, Eq. (12).
The second is a proof that the two definitions of chaos in Definition 8 are
inequivalent. Define
\begin{equation}
F(x)=Mx+w-(\frac{a}{2\pi })\sin 2\pi x-(\frac{b}{4\pi })\sin (4\pi x).
\end{equation}
\newline
\newline
\textbf{Proposition 5:} If the following conditions are satisfied then the
rotation number of $F(x)$ in Eq. (20) is 1/2
\begin{eqnarray}
M+w &>&1+(\frac{a}{2\pi })+(\frac{b}{4\pi }),\;w+(\frac{a}{2\pi })+(\frac{b}{%
4\pi })<1,  \nonumber \\
a+b &<&M<1, \\
w &>&\theta ,\;\;M+w<1+\theta .  \nonumber
\end{eqnarray}
\newline
\newline
\textbf{Proof.} Equation (20) implies
\[
Mx+w+\frac{a}{2\pi }+\frac{b}{4\pi }\geq F(x)\geq Mx+w-\frac{a}{2\pi }-\frac{%
b}{4\pi },
\]
thus
\[
\frac{1+(a/2\pi )+(b/4\pi )-w}{M}\geq \theta \geq \frac{1-(a/2\pi )-(b/4\pi
)-w}{M},
\]
where $\theta $ is defined in Eq. (19). The first two conditions in Eq. (21)
guarantee that $0<\theta <1.$

The condition $a+b<M$ implies that $F(x)$ satisfies FII. The condition $M<1$
implies that $F(0)>F(1)$. This guarantees that the discontinuity of $F$ is
unique. Thus $F$ satisfies FI-FIV hence by Keener's Lemma 3.1 [Keener 1980]
the rotation number of $f$ is a single point.

To prove that the rotation number equal to $1/2$ use the third set of
conditions in Eq. (21) which implies
\[
F(0,\theta )\subset (\theta ,1),\;F(\theta ,1)\subset (0,\theta )\Rightarrow
\;F^{2}(0,\theta )\subset (0,\theta ),
\]
hence [Holmgren 1996] $F^{2}$ has a fixed point in $(0,\theta )$ i.e. the
rotation number equal to $1/2$. $\blacksquare $\newline

To see that the conditions in Eq. (21) are consistent let
\[
M=0.8,\;w=0.6,\;\theta =0.5,\;a=b=0.1.
\]
Now the following proposition can be proved. Let
\begin{equation}
F=Mx+w\mathrm{\;mod\;}1.
\end{equation}
\newline
\newline
\textbf{Proposition 6:} The two definitions of chaos (c.f. Definition 8) in
the sense of Keener and the standard sense are inequivalent.\newline
\newline
\textbf{Proof.} Consider the map (20) with the following conditions
\begin{equation}
2-w>M>1,\;0<w<1.
\end{equation}
The first condition in Eq. (23) implies that $F$ satisfies FI, FII, FIII, FV
(overlapping condition $F(1)>F(0)$). The second condition implies that $x=$ $%
(1-w)/(M-1)\;$is a fixed point and the third one implies $0<\theta =(1-w)/M.$
Since $x=(1-w)/(M-1)$ is a fixed point of $F$ it belongs to $I$. Using Eq.
(23), one gets
\[
F(1)=M+w-1<\frac{1-w}{M-1},
\]
then $I$ is not subset of $[F(0),F(1)]$. Moreover since $x=0$ is not a fixed
point of $F$, then $I\neq [0,1]$. Thus by Theorem 1 the rotation number of $%
f $ is a single point. Yet Eq. (22) implies that Lyapunov exponent of $F$ is
positive. This completes the proof. $\blacksquare $

\section{Ito-Glass coupled map lattice}

Ito and Glass [Ito and Glass 1992] have proposed a coupled map lattice which
is more suitable for modelling the heart. The equations are given by
\begin{equation}
x_{i}^{t+1}=-g(x_{i}^{t})+l[\sum_{j=1}^{i-1}f(x_{j}^{t+1})+%
\sum_{j=i}^{n}f(x_{j}^{t})],
\end{equation}
where $i=1,2,...,n$. The homogeneous equilibrium solution is given by
\begin{equation}
x=-g(x)+nlf(x).
\end{equation}
Define $a$ and $c$ by
\begin{equation}
a=l\frac{\mathrm{d}f(x)}{\mathrm{d}x},\;c=\frac{\mathrm{d}g(x)}{\mathrm{d}x}.
\end{equation}
Using inhomogeneous analysis they were able to show that the characteristic
polynomial for the equilibrium solution Eq. (25) is
\begin{equation}
\frac{(-1)^{n}\{\lambda (\lambda +c)^{n}-[(1+a)\lambda -a+c]^{n}\}}{\lambda
-1}=0.
\end{equation}

We begin by studying the simplified system:
\begin{equation}
\theta _{i}^{t+1}=\sum_{k=1}^{i-1}f(\theta
_{k}^{t+1})+\sum_{k=i}^{n}f(\theta _{k}^{t}).
\end{equation}
The homogeneous equilibrium solution is given by
\begin{equation}
\theta =nf(\theta ).
\end{equation}
Using Eq. (27) the characteristic polynomial for the solution (29) is
\[
\frac{\lambda ^{n+1}-[(1+a)\lambda -a]^{n}}{\lambda -1}=0,
\]
where $a=f^{\prime }(\theta ),\;\theta $ is given by Eq. (29). Using
l'Hopital rule it can be written as
\begin{equation}
\lambda ^{n}-[n/(n+1)](1+a)[(1+a)\lambda -a]^{n-1}=0\;\;\;
\end{equation}

Since we are interested in modelling the heart we are interested in the case
$n$ is large. Thus we have:\newline
\newline
\textbf{Proposition 7:}

\begin{enumerate}
\item  The equilibrium solution, Eq. (29) is unstable if any one of the
following conditions are satisfied:

\begin{equation}
\begin{array}{l}
(\mathrm{i})a>0,\;(1+a)^{n+1}\gg a(n+1)/[n(1+a)]. \\
(\mathrm{ii})a<-2,\;n\mathrm{\;is\;sufficiently\;large}. \\
(\mathrm{iii})-2<a<-1,\;n\;\mathrm{is\;odd\;and\;sufficiently\;large}.
\end{array}
\end{equation}

\item  There are no real roots for Eq. (30) satisfying $\left| \lambda
\right| >1$ if $-1\leq a\leq 0,$ and $n$ is sufficiently large.

\item  Define
\begin{equation}
\alpha _{0}=1,\;\alpha _{k+1}=\frac{(-1)^{k+1}n(n-1)!}{(n+1)k!(n-k-1)!}%
a^{k}(1+a)^{n-k},\;k=0,1,2,...,n-1,
\end{equation}

then if $-1\leq a\leq 0$ and $n$ is sufficiently large such that $\alpha
_{k+1}\ll 1\;\forall k=0,1,2,...,n-1,$ then the equilibrium solution (29) is
stable.
\end{enumerate}

\textbf{Proof.} We will prove (i) by showing the existence of a root $%
\lambda >1$ for Eq. (30). If there is a root $\lambda \gg a/(a+1),$ then Eq.
(30) can be approximated by
\[
\lambda ^{n}\simeq \frac{n}{n+1}(1+a)^{n}\lambda ^{n-1}\Rightarrow
\;\;\lambda \simeq \frac{n}{n+1}(1+a)^{n}>1\;\mathrm{for}\;n \gg 1.
\]
The same argument is valid for $a<-2$ and $n$ sufficiently large hence (ii)
is proved.

To prove (iii) we will show that there is a positive integer $k>1$ such that
the LHS of Eq. (30) is positive at $\lambda =-k+1$ and negative at $\lambda
=-k.$ Since $n$ is assumed odd the first term in Eq. (30) is always negative
for $\lambda <0$ while the second term is positive (notice that $1+a<0$).

For $n$ sufficiently large the positivity at $\lambda =-k+1\;$implies\
\[
k\left| a\right| >2(k-1),
\]
while the negativity at $\lambda =-k\;$implies\ $(k+1)\left| a\right| <2k.$
Thus whenever $2(k-1)/k<\left| a\right| <2k/(k+1)$ there is a root $\lambda
\in [-k,-k+1].$

To prove the second part of the proposition assume $-1\leq a\leq 0.$ The
proof is by direct substitution for $a=0$, $a=-1$. For these two values the
homogeneous solution (29) is stable.

For\ $-1<a<0\;$it is direct to see that, in this case, Eq. (30) does not
admit a negative solution. Thus the only remaining possibility is a $\lambda
>1$ solution. To show that this is not possible, define $f(\lambda )$ to be
the right hand side of Eq. (30) and $f^{k}(\lambda )$ to be the $k$-th
derivative of $f$ with respect to $\lambda $. Then it is direct to see that
\[
f^{k}(1)=n(n-1)(n-2)...(n-k+1)\left[ 1-\frac{(n-k)(1+a)^{k+1}}{n+1}\right]
>0\;\forall k=0,1,2,....
\]
Consequently, using Taylor series we have $f(\lambda )>0\;\forall \lambda
\geq 1$. Therefore there are no real roots $\lambda >1$ for Eq. (30) in the
case $-1\leq a\leq 0.$

We prove the third part of the proposition using Jury test [Murray 2002] so
we define
\begin{eqnarray*}
b_{n-k} &=&\alpha _{k}-\alpha _{n}\alpha _{n-k},\;0\leq k\leq n-1, \\
\;c_{n-k} &=&b_{n}b_{n-k}-b_{1}b_{k+1},\;0\leq k\leq n-2, \\
d_{n-k} &=&c_{n}c_{n-k}-c_{2}c_{k+2},\;0\leq k\leq n-3, \\
&&... \\
q_{n-k} &=&p_{n}p_{n-k}-p_{n-3}p_{k+n-3},\;0\leq k\leq 2.
\end{eqnarray*}
The conditions we assume in part (3) of the proposition implies
\[
b_{n-k}\simeq \alpha _{k},\,\,c_{n-k}\simeq \alpha _{k},\;d_{n-k}\simeq
\alpha _{k},...,\;\,q_{n-k}\simeq \alpha _{k},
\]
hence
\begin{equation}
\left| \alpha _{n}\right| \ll 1,\left| b_{1}\right| \ll \left| b_{n}\right|
,\left| c_{2}\right| \ll \left| c_{n}\right| ,\left| d_{3}\right| \ll \left|
d_{n}\right| ,...,\left| q_{n-2}\right| \ll \left| q_{n}\right| .
\end{equation}
Since
\begin{eqnarray*}
f(1) &=&1-\frac{n(1+a)}{n+1}\Rightarrow f(1)>0, \\
(-1)^{n}f(-1) &=&1+\frac{n(1+a)}{n+1}(2a+1)^{n-1}>0,\;\forall -1<a<0,
\end{eqnarray*}
then the conditions of Jury test are satisfied. $\blacksquare$\newline

Applying Proposition 7 to the circle map
\begin{equation}
f(x)=x+w+\frac{a}{2\pi }\sin (2\pi x),
\end{equation}
one gets that if $k<1$ then the homogeneous solution for the Ito-Glass CML
for the circle map (34) is unstable if (31) is valid.

We introduce the following local version of Ito-Glass coupled map lattice
(IGCML)
\begin{equation}
x_{i}^{t+1}=(1-D)f(x_{i}^{t})\;+\frac{D}{2}\left[
f(x_{i-1}^{t+1})\;+f(x_{i+1}^{t})\right] ,\;i=1,2,...,n\;\mathrm{and}\;0\leq
D<1.\;\;\;
\end{equation}
The homogeneous equilibrium solution is given by
\begin{equation}
x=f(x).
\end{equation}
\newline
\newline
\textbf{Proposition 8:} The equilibrium solution (36) is stable if $\left|
f^{\prime }(x)\right| <1$.\newline
\newline
\textbf{Proof.} Assume
\[
x_{i}^{t}=x+\varepsilon _{i}^{t},\;\left| \varepsilon _{i}^{0}\right| <\eta
\;\forall i=1,2,...,n.
\]
Using the conditions $0\leq D<1,\;\left| f^{\prime }(x)\right| <1\;$then
\[
\varepsilon _{1}^{1}=(1-D)f^{\prime }(x)\varepsilon _{1}^{0}+\frac{D}{2}%
f^{\prime }(x)\varepsilon _{2}^{0}\;\;\Rightarrow \left| \varepsilon
_{1}^{1}\right| <\eta ,
\]
hence
\[
\varepsilon _{2}^{1}=(1-D)f^{\prime }(x)\varepsilon _{2}^{0}+\frac{D}{2}%
f^{\prime }(x)(\varepsilon _{2}^{0}+\varepsilon _{1}^{1})\;\Rightarrow
\;\left| \varepsilon _{2}^{1}\right| <\eta ,
\]
and so on. Thus $\left| \varepsilon _{i}^{t}\right| <\eta \;\;\forall
i=1,2,...,n\;\mathrm{and}\;\forall t>0.$ This completes the proof. $%
\blacksquare$\newline

We introduce another form of global Ito-Glass coupled map lattice which is
significantly closer to those of Kuramoto, Winfree etc.. [Ahmed and Hegazi
2002] is
\begin{eqnarray}
x_{i}^{t+1} &=&x_{i}^{t}+w_{i}+\frac{k}{n}\left[
\sum_{j=1}^{i-1}f(x_{j}^{t+1})\;+\;\sum_{j=i}^{n}f(x_{j}^{t})\right] , \\
0 &\leq &w_{i}<1\;\forall i=1,2,...,n\;,\;1>k\geq 0.\;\;\;\;  \nonumber
\end{eqnarray}
The equilibrium solution is given by
\begin{equation}
w+kf(x)=0.
\end{equation}
\newline
\newline
\textbf{Definition 9:} A dynamical system is persistent if $%
x_{i}^{0}>0\;\forall i=1,2,...,n$ implies $\lim_{t\rightarrow \infty }\inf
(x_{i}^{t})>0\;,\forall i=1,2,...,n.$\newline
\newline
\textbf{Proposition 9:}

\begin{enumerate}
\item[(a)]  For sufficiently large $n,$ the solution (38) is stable if $%
f^{\prime }(x)<0.$

\item[(b)]  The system (37) is persistent if $f^{\prime }(0)>0.$
\end{enumerate}

\textbf{Proof. }Applying l'Hopital rule to Eq. (27), one gets that the
characteristic polynomial for the homogeneous solution of the general
Ito-Glass CML is
\begin{equation}
(\lambda +c)^{n}+n\lambda (\lambda +c)^{n-1}-n(1+a)[(1+a)\lambda
-a+c]^{n-1}=0.
\end{equation}
The system (37) corresponds to
\[
c=-1,\;a=\frac{k}{n}f^{\prime }(x).
\]
Substituting in Eq. (39), one gets $\lambda =1\;$or
\[
\lambda =\frac{1}{n+1}+\frac{n}{n+1}\left[ 1+\frac{k}{n}f^{\prime
}(x)\right] ^{n},
\]
for $n$ sufficiently large
\[
\lambda \simeq \exp (kf^{\prime }(x)),
\]
which proves (a).

To prove (b): Assume $1\gg x_{i}^{t}>0$ for some $t$ and $\forall
i=1,2,...,n $ and use Eq. (37) hence
\begin{eqnarray*}
x_{i}^{t+1} &\simeq &x_{i}^{t}+w_{i}+f^{\prime }(0)(\frac{k}{n})\left[
\sum_{j=1}^{i-1}x_{j}^{t+1}\;+\;\sum_{j=i}^{n}x_{j}^{t}\right] \Rightarrow ,
\\
x_{1}^{t+1} &\simeq &x_{1}^{t}+w_{1}+f^{\prime }(0)(\frac{k}{n}%
)x_{2}^{t}\;>x_{1}^{t}, \\
x_{2}^{t+1} &\simeq &x_{2}^{t}+w_{2}+f^{\prime }(0)(\frac{k}{n})\left[
x_{1}^{t+1}\;+\;\sum_{j=i}^{n}x_{3}^{t}\right] >x_{2}^{t},
\end{eqnarray*}
and so on. This completes the proof of the proposition. $\blacksquare$%
\newline

Another form of Ito-Glass coupled map lattice is
\begin{eqnarray}
x_{i}^{t+1} &=&(1-k)f(x_{i}^{t})+\frac{k}{n-1}\left[
\sum_{j=1}^{i-1}f(x_{j}^{t+1})\;+\;\sum_{j=i}^{n}f(x_{j}^{t})\right] , \\
0 &\leq &w_{i}<1\;\forall i=1,2,...,n\;,\;1>k\geq 0.  \nonumber
\end{eqnarray}
The homogeneous equilibrium solution is given by $x=f(x).$ If $k\ll 1$ then
it is direct to see that the equilibrium solution is stable if $\left|
f^{\prime }(x)\right| <1.$ For general $k$ ($1>k\geq 0$) and sufficiently
large $n$ one has\newline
\newline
\textbf{Proposition 10:} For general $k$ ($1>k\geq 0$) and sufficiently
large $n$ the roots of the characteristic polynomial of Eq. (40) are
\begin{equation}
\lambda =-c,\;\mathrm{and}\;\;\lambda \simeq \exp \left[ -kf^{\prime }(x)%
\frac{1-\lambda }{\lambda +c}\right] ,\;c=(-1+k)f^{\prime }(x).
\end{equation}
Moreover, if $-1<f^{\prime }(x)<0$ then the equilibrium solution $x=f(x)$ is
stable.\newline
\newline
\textbf{Proof. }The proof of Eq. (41) is similar to that of part (a) in
Proposition 9.

Now assume that $-1<f^{\prime }(x)<0,$ then $\left| c\right| <1.$ Thus
instability may arise only from the transcendental equation in Eq. (41). If $%
\lambda >1,$ then $-kf^{\prime }(x)(1-\lambda )/(\lambda +c)\;$is negative
hence $\exp [-kf^{\prime }(x)(1-\lambda )/(\lambda +c)]<1$, contradiction.
Since $\exp [-kf^{\prime }(x)(1-\lambda )/(\lambda +c)]\geq 0,$ then $%
\lambda <-1$ is not possible. $\blacksquare $

\section{Conclusions}

Some generalizations of the circle map are given. The two definitions of
chaos (in the sense of Keener and the standard one) are found to be
inequivalent. Ito-Glass coupled map lattice is more suitable for modelling
the heart. Many forms of Ito-Glass coupled map lattice are introduced and
studied. The conditions for the stability of its equilibrium solutions are
derived.

\section*{Acknowledgments}

We Thank B. Duggal and the referees for their helpful comments. One of the
authors (A. S. Hegazi) acknowledge the financial support of Mansoura
University grant.

\end{document}